\documentclass[12pt]{article}       
\usepackage{multicol}

\renewcommand{\baselinestretch}{1.} 

\begin{document}

\thispagestyle{empty}

{\large
\parskip 10 pt

\hfill{}

\vspace{2 in}
\centerline{\LARGE\bf Life Cycles of Magnetic Fields in Stellar Evolution}
\vspace{1cm}
\centerline{\bf A White Paper Submitted to the Astro-2010 Decadal Survey}

\vspace{1.cm} 
\hfill{}\\
\vspace{5 pt} 
Dmitri A.~Uzdensky, Princeton University; \ uzdensky@astro.princeton.edu \\
\vspace{5 pt} 
Jonathan Arons,\ University of California, Berkeley \\
\vspace{5 pt} 
Steven A.~Balbus,\ \'Ecole Normale Sup\'erieure\\
\vspace{5 pt} 
Eric G.~Blackman,\ University of Rochester \\
\vspace{5 pt} 
Jeremy Goodman,\ Princeton University \\
\vspace{5 pt} 
Mikhail V.~Medvedev,\ University of Kansas \\
\vspace{5 pt} 
Anatoly Spitkovsky,\ Princeton University \\
\vspace{5 pt} 
James M.~Stone,\ Princeton University 

\vspace{1cm} 
{\it Endorsed by the Topical Group on Plasma Astrophysics 
of the American Physical Society}


\vspace{1 cm}
\underline{Science Frontier Panel:} 
\quad Stars and Stellar Evolution (SSE)
}
\parskip 0 pt



%
%

\newpage
\setcounter{page}{1}

\centerline{\bf\Large Life Cycles of Magnetic Fields in Stellar Evolution}

\vspace{15 pt}
\centerline{\bf\large D.~Uzdensky, J.~Arons, S.~Balbus, E.~Blackman,
J.~Goodman,}
\vspace{5 pt}
\centerline{\bf\large M.~Medvedev, A.~Spitkovsky, \& J.~Stone}

\vspace{10 pt}
\centerline{\large A white paper submitted to the Astro-2010 Decadal Survey}

\vspace{10 pt}
{\it Endorsed by the Topical Group on Plasma Astrophysics 
of the American Physical Society.}


\section{Introduction}

Understanding the structure and evolution of the stars is 
a great triumph of 20th century Astrophysics. 
But this has also revealed a plethora of unsolved problems 
associated  with the magnetization of stellar matter.  
These include magnetic mediation of star formation, 
rotational structure and evolution of main sequence stars, 
the origin of stellar magnetism itself, stellar mass loss, 
coronal activity, etc. The death of stars and their after-life 
as compact objects presents a variety of additional puzzles 
whose answers are also illuminating for understanding similar 
phenomena in galaxies harboring supermassive black holes.
Questions of wide interest for the coming decade are those related 
to the death of massive stars ---  supernovae (SNe) and the remnants 
they leave behind (neutron stars and black holes), including the relativistic 
outflows 
from neutron stars and the disks around black holes, and the impulsive 
jets inferred in~GRBs, both those associated with SNe and those perhaps 
powered by merging neutron stars.  
All of these phenomena are directly or indirectly tied to the magnetic 
fields embedded in the stellar material. Thus all are illuminated by 
the incorporation of plasma physics into astrophysics, the realm of  
{\it Plasma Astrophysics}.

%

Prospects for further progress in Plasma Astrophysics in the next decade 
are very bright, and are due to recent and anticipated advances in theory, 
simulations, and the advent of laboratory plasma-astrophysics experiments.

An over-arching theme in plasma physics as applied 
to astrophysical objects is the {\it life cycle of magnetic fields}: \\
(1) Their origin: How are they produced and amplified? \\
(2) Their active life: how do they interact with the plasma
and affect the dynamic behavior of various astrophysical systems? \\
(3) and finally, their destruction: how are they destroyed (dissipated)
and what is the fate and observable signatures of the released magnetic
energy? 

These themes come up repeatedly in various astrophysical 
situations and have direct observational implications. 
We also note that these life-cycle processes occur on 
two physically different scales: macroscopic/MHD scales 
(MHD dynamo, MRI, collimation and acceleration of jets/winds, 
magnetic braking); and microscopic/kinetic plasma scales 
(collisionless shocks, reconnection, plasma micro-turbulence, 
non-thermal particle acceleration, radiation). 

This white paper will highlight the following topics: \\
(1) The Origin of Magnetic Fields and Magnetic Activity in Stars \\
(2) Accretion Disks \\
(3) Stellar explosions: SNe and GRBs. 

\section{Stellar Magnetic Fields and Magnetic Activity}

Magnetic fields are evident in all phases of stellar evolution. 
Not only are they directly measured in main sequence stars and 
white dwarfs, but their presence is also revealed indirectly 
through active stellar coronae, outflows, and the evolution of 
stellar rotation (e.g., in neutron stars). This raises the following 
broad unifying issues that we discuss in this section:
the origin of magnetic fields in stars including 
the Sun; their dynamical role in stellar evolution; 
and the stellar coronal and outflow activity powered 
by magnetic dissipation.



\subsection{MHD Dynamo and the Origin of Stellar Magnetic Fields}

Global reversals of the solar magnetic field imply in situ field 
regeneration by a dynamo, and rule out a fossil field origin.  
However, understanding what the minimum ingredients for a dynamo are,  
how a dynamo grows fields of large enough scale to survive into the corona, 
and  linking this field growth directly to stellar coronal activity comprise 
an enterprise of research.
The subject demands  understanding the interplay between convection, 
MHD turbulence, differential rotation, and  stratification. The scale 
separation and the complex mutual interdependence of these processes 
requires a symbiosis between semi-analytic approaches parametrizing 
nonlinear features and carefully designed numerical simulations.
Recent theoretical developments and evolving numerical tools bode 
well for substantial progress in this area in the next decade.
Furthermore, advances in stellar dynamo will also impact our 
understanding of the origin of galactic, and accretion-disk, 
magnetic fields.

Dynamo action can be classified into two categories:
(1) Small-scale dynamo in which magnetic fields 
are amplified at and below the driving scale 
and 
(2) Large-scale dynamo (LSD) in which an initially weak 
large-scale field is amplified with a sustained flux 
on time and/or spatial scales significantly larger than 
that of the turbulence.  
Large-scale dynamos are of particular interest in astrophysics 
since they are needed to explain the solar magnetic field and 
probably the origin of large-scale fields of any rotating convective 
star (including proto-neutron stars).
The key unifying quantity in LSDs is a magnetic-field-aligned mean~EMF.

A fundamental impasse with earlier LSD theories 
is that they were kinematic, and hence could not predict the saturation 
level of the large-scale field. 
Recently, however, new principles and their numerical tests have emerged, 
opening the door to progress: a unifying role of the magnetic helicity 
in LSDs has been recognized, with large-scale growth of one sign, 
and small-scale dissipation of the opposite sign. 
Coupling the dynamical evolution of magnetic helicity into the
LSD equations has led to a semi-analytic nonlinear dynamo model 
for the large-scale field saturation~\cite{Blackman_Field-2002}, 
confirmed by numerical simulations~\cite{Brandenburg-2001}. Studies 
using open boundaries, convection, and shear, have also lent support 
to the paradigm that tracking the flow of the large- and small-scale 
magnetic helicity has predictive power~\cite{Blackman_Field-2000, 
Vishniac_Cho-2001,Sur-2007}.
Incorporating the principle of magnetic helicity evolution  
into more realistic global dynamo models and developing a 
practical ``textbook'' theory is a challenge for the next decade.


The MHD dynamo problem is intrinsically related to MHD turbulence,
and the last decade has seen great progress in our theoretical 
understanding of this challenging subject~\cite{Goldreich_Sridhar-1995, 
Goldreich_Sridhar-1997,Politano_Pouquet-1998,Boldyrev-2005}.
In a realistic turbulent system, small- and large-scale field 
amplification is often contemporaneous.
While in the past LSD theory has often focused on field growth 
at the largest scale, turbulence approaches focus on the overall 
spectrum \cite{Brandenburg-2001,Boldyrev_etal-2005}, and so the 
term ``LSD'' means the generation of magnetic fields on any scale 
larger than the turbulent scale. 



Regarding future progress, recent numerical studies indicate that proper 
simulations of MHD dynamo (and MHD turbulence in general) must include 
small but finite {\it explicit} resistivity and viscosity corresponding 
to rather large Reynolds and magnetic Reynolds numbers, at least $10^3-10^4$.
Although the required computing power has not been easily available 
in the past, it is now becoming common place. Thus, the application 
and further development of analytical theory combined with powerful 
numerical simulations gives us a unique opportunity to solve this 
important problem.

Finally,  new liquid-metal and plasma laboratory experiments 
(existing and proposed) are underway to study flow-driven MHD 
dynamos under controlled  conditions.
Although these experiments cannot address the nonlinear saturation 
of~LSDs, they can test the key basic principles of the kinematic 
dynamo, which so far have been assumed or tested only numerically. 
In addition, dynamos in magnetic-confinement plasma devices such as 
Spheromaks and RFPs, which describe the relaxation of strongly-magnetized 
systems  to large scales in response to small-scale magnetic helicity 
injection, are  directly related to field opening and relaxation 
in coronal plasmas. This analogy should be exploited.


\subsection{Differential Stellar Rotation}

Convectively driven dynamos in planetary cores and in the outer layers
of late-type stars (e.g., the Sun) are associated with magnetized, 
{\it differentially rotating} flows. The origin of these differentially 
rotating flows is not well understood, but it is known that their stability 
is profoundly affected by even very weak magnetic fields, as is beautifully 
exemplified by the magneto-rotational instability (MRI) in accretion disks.
An important clue is that recent helioseismology results and the vorticity 
equation together imply a near confluence of surfaces of constant angular 
velocity and constant entropy throughout the bulk of the solar convection 
zone.  This, as has been recently noted~\cite{Balbus-2008}, is just the
expected configuration of a weakly magnetized, differentially rotating,
convecting gas that is marginally (un)stable to ``magneto-baroclinic''
modes enabled by a weak magnetic field.  Clearly, the consequences of MHD
processes in rotating convective systems has yet to be fully elucidated.


\subsection{Coronal Activity}

Most of the stellar dynamo studies described above naturally focus on 
flow-dominated amplification of the magnetic field that takes place in
the interiors of stars. However, the magnetic activity that we actually
observe takes place in the stellar coronae.
It is thus important to prioritize research that can make the connection 
between the fields produced inside a star and what emerges (due to magnetic
buoyancy) into the corona. Similar arguments also apply to the coronae of 
accretion disks. Once in a star's (or a disk's) corona, the magnetic field
structures are still evolving, in part towards the largest possible scales.
Thus, the analogy between stellar dynamo, coronal fields, and the large-scale 
fields of coronal holes, and the accretion disc dynamos and the large-scale 
fields of jets should be pursued.

Whereas the dynamo regions where magnetic fields are generated 
are relatively dense and cold, the coronae of stars and accretion 
disks, where the field destruction actually takes place, are 
hot and rarefied. Therefore, whereas resistive MHD is probably 
a good description in the former environment, it may not be valid 
in the latter. Thus, the life of magnetic fields is not limited to 
single-fluid MHD, but goes well beyond, into plasma physics proper, 
requiring a two-fluid or even a kinetic description. 
In particular, non-thermal particles and plasma microturbulence are 
likely to play an active role in the low-density collisionless or 
marginally collisionless coronal environments.

In this regard, we would like to stress the importance of recent 
advances in magnetic reconnection research, including theory, numerical 
simulations, and lab experiments. 
The emerging understanding of the role of plasma collisionality 
and of collisionless plasma processes is only now starting to 
percolate and to be applied in today's astrophysics, namely,
to astrophysical coronae~\cite{Uzdensky-2007}. 
This work should continue since it may finally lead to understanding 
the principles governing these coronae and their emission.








\section{Accretion disks} 

Accretion disks are essential to the formation, evolution, and observable
radiation of many classes of stars and compact objects, including protostars
and protoplanets, CVs, XRBs, and supermassive black holes.  The process of 
accretion requires that mass be separated from angular momentum; understanding
this angular momentum transport (AMT) has been a longstanding problem.
There are good theoretical grounds for believing that AMT occurs via
MHD turbulence driven by the MRI, though magnetized winds and jets are 
also candidate mechanisms. Magnetic and plasma processes clearly lie 
at the heart of accretion-disk dynamics.  Recent numerical research 
indicates that the intensity of MRI turbulence may depend on many 
factors, e.g., under certain circumstances, on the microphysical 
dissipation coefficients~\cite{Fromang-2007}.  Determining when 
MRI-turbulence in accretion disks is a self-sustaining dynamo is 
a very active area of research.  
A proper understanding of it will require very high-resolution 
3D simulations with large values and ranges of Reynolds and magnetic 
Reynolds numbers. This should be feasible in the next decade.

One of the richest and most challenging applications of AMT studies is 
to cool disks, namely, protostellar disks and the outer parts of AGN disks.  
These media are characterized by small ionization fraction, high resistivity, 
and poor gas--magnetic field coupling.  
The ionization is governed by a poorly understood balance involving 
nonthermal ionization sources, dust physics, and molecular chemistry.
The dominant ionization sources in protostellar disks are the coronal 
X-rays from the central star and Galactic cosmic rays~\cite{Gammie-1996,
Glassgold_etal-1997}; 
in AGN, it is probably the nonthermal hard X-ray emission from the inner 
disk.  In all cases, understanding the nature of a magnetized hot plasma 
is central.



The difficult, longstanding problems raised above belong to 
the domain of MHD and kinetic plasma physics. The effort 
required to address is justified because the prospects for 
progress over the next ten years are good, and because an
understanding of these processes is critical to problems at 
the forefront of astrophysics.  Consider but two very important 
problems in modern astronomy: planet formation, and the joint evolution 
of galaxies and their central black holes. To understand the emergence 
of planets in a protoplanetary disk, for example, one has to know how 
the intensity of turbulence depends upon macroscopic disk parameters 
(e.g., density, temperature, and irradiation) and how the turbulence 
in turn governs the rates at which dust settles, planetary cores grow, 
and planetary orbits migrate.  In AGN, the microscopic exchange of 
energy among electrons, ions, and magnetic fields not only affects 
how we deduce accretion rates from observations, but also is critical 
to the balance between accretion and ejection, hence the feedback on 
the host galaxy.

What are the prospects for progress?  
First, MRI-driven turbulence, the most plausible fundamental reason 
why accretion in disks occurs, has now become a mature field of research.
Although technically difficult, liquid-metal experimental studies of the MRI 
have already achieved greater hydrodynamical Reynolds numbers than are 
accessible to the most powerful numerical simulations, thereby placing
rigorous limits on any non-magnetic turbulent transport in rotating
flow~\cite{Ji_etal-2006}.  With plausible
increases in scale, these experiments could produce MRI turbulence in
the important low magnetic Prandtl number regime.
On the observational front, we expect that molecular spectroscopy and
direct imaging of protostellar disks via ALMA and SOFIA will provide
valuable results, constraining or determining disk densities, temperatures, 
ionization fractions, and chemistry. The prospects for a similar progress 
in X-ray studies of black-hole disks are less certain. However, there is 
a growing confidence among plasma physicists, based on numerical simulations 
and lab experiments, that the mechanisms for fast magnetic reconnection 
(presumably powering accretion-disk coronae) in collisionless plasmas 
have been identified.

The focus of accretion disk studies is now shifting from the classical
AMT problem towards small-scale  dissipation, detailed thermodynamics, 
and radiation in accretion flows.  Important questions that will dominate 
this area in the next decade include the following: How and where is the 
turbulent kinetic and magnetic energy dissipated into heat to produce the 
observed emission? Does dissipation via reconnection occur in the collisional
(resistive-MHD) or collisionless regime?  What causes spectral-state
transitions in accreting black holes?  What fraction of the accretion 
energy is dissipated locally in the disk and what fraction in an overlying 
corona?  How much energy goes into an outflow?  How does the small-scale MHD 
turbulence interact with a large-scale external magnetic field?  
How is material transferred from the inner edge of a disk to a central, 
possibly magnetized, star?

In brief: the focus of accretion disk studies 
is shifting toward the physics of plasmas.



\section{Gamma-Ray Bursts}

\subsection{GRB/SN Central Engines}

Long Gamma-Ray Bursts (GRBs), perhaps the most powerful and enigmatic
explosions in the Universe, are still only poorly understood. Are they 
associated with supernovae (SNe) or failed~SNe? Are they powered by a 
hyper-accreting black hole, as in the collapsar model, perhaps via the 
Blandford-Znajek mechanism, or by a rapidly-rotating magnetar? What are 
their characteristic neutrino- and gravitational-wave signatures? 
How do their relativistic jets interact with the surrounding stellar 
envelope and collimate? 
Are they kinetic-energy-dominated (as in the traditional fireball/internal 
shock picture) or Poynting-flux dominated? How can one observationally 
distinguish between these alternatives?

In the past few years, traditional, purely neutrino-driven models of 
SNe and GRB have come into doubt. This has led to a growth of interest 
in magnetically-driven core-collapse explosions of massive stars, in which 
the central engine's rotational energy is extracted magnetically to power 
the explosion (e.g., via a magnetic-tower-like mechanism). In fact, this 
general mechanism is applicable to a broad class of asymmetric stellar 
explosions/outflows, e.g., to planetary nebulae --- the end states of 
low-mass stars~\cite{Balick_Frank-2002}. 

Future theoretical progress in understanding the central engines 
of SNe and especially GRBs will require a better grasp of basic 
plasma physics relevant to this extreme, high-energy-density 
environment. Thus, assessing the dynamical role of magnetic fields 
makes the task of understanding the classical MHD processes 
(dynamo, MRI, magnetic towers, jet acceleration and collimation, 
Parker and kink instabilities, reconnection) in the context of 
collapsing stars, imperative. Significant numerical progress in
this area is expected in the next decade due to the advent of 
general-relativistic MHD codes employing advanced numerical algorithms
and incorporating realistic microphysics (including neutrino processes), 
combined with the ever-increasing computer power.


\subsection{Collisionless Shocks, Radiation, and ``a GRB in a Lab"} 
\label{GRB+lab}

Even if initially the jet is magnetically dominated, 
in the end one still has to understand how the jet 
energy is eventually converted to the observable radiation.
There are, in principle, two main ways of doing this. 
One is the classical {\it internal shock model}: 
slow, macroscopic magneto-centrifugal acceleration 
gradually converts the Poynting flux to the bulk kinetic 
energy of the relativistic flow, which is later dissipated 
in microscopically-thin collisionless shocks producing 
nonthermal electrons that subsequently radiate away their 
energy. In the other case, a magnetically dominated jet 
may undergo direct, microscopic dissipation via reconnection 
in current sheets produced by MHD instabilities or by initial 
irregularities in the jet; non-thermal particles accelerated 
in the reconnection region then produce the observed emission.
Both of these scenarios involve collective plasma processes 
at microphysical scales (plasma microturbulence) coupled with 
nonthermal particle acceleration and radiation. We shall now 
discuss the first (more widely accepted) scenario in more detail.


GRBs are the most extreme end of cosmic blast waves, and their 
detailed physical understanding has to be based on the merger 
and interpenetration of astrophysics and plasma physics 
(namely, high-energy-density and relativistic plasma physics). 
The past decade illustrated the fruitfulness of such a process, 
as we went from the first theoretical realization of the importance 
of the Weibel instability (a kinetic anisotropy-driven plasma instability) 
for the formation of GRB collisionless shocks~\cite{Medvedev_Loeb-1999}
to breathtaking kinetic plasma simulations~\cite{Silva_etal-2004, 
Frederiksen_etal-2004,Spitkovsky-2008}, 
and we are now about to produce 
Weibel-mediated shocks with nearly realistic (astrophysical!) plasma 
conditions on Earth with Petawatt-scale lasers. 
The traditional picture of GRBs holds that they launch ultra-relativistic 
(with Lorentz factors $\sim 10^2$) outflows that produce relativistic 
collisionless shocks both inside the outflow and 
at the ambient medium interface. 
At such shocks, the plasma (unless strongly magnetized) breaks up into 
an array of small-scale current filaments via the Weibel instability, 
generating strong magnetic fields. The resulting nonlinear plasma 
turbulence scatters particles, effectively introducing collisions 
into the otherwise collisionless plasma. Moreover, radiation emitted 
by the shock-accelerated electrons in such sub-Larmor-scale magnetic 
fields is markedly non-synchrotron. A radiating electron never completes 
its gyro-orbit but, instead, jitters around its nearly straight trajectory, 
possibly accounting for the observed benchmark harder-than-synchrotron 
``jitter'' radiation spectrum~\cite{Medvedev-2000}. 

While the general picture seems to be understood, many questions 
remain open. What is the dynamics of the nonlinear Weibel turbulence 
in the upstream precursor and the downstream post-shock plasmas? 
How does this turbulence heat the electrons? 
What role do the Weibel fields play in Cosmic Ray (CR) acceleration? 
What feedback do the CRs exert on the shock by modifying the upstream 
medium? How is radiation produced in Weibel-mediated shocks?
All these questions are crucial for our understanding of GRBs. 

Most of the present-day results come from extensive PIC simulations. 
This line of research will grow as Petaflop-scale computers become
available, which will enable us to model 3D ion-electron collisionless 
shocks, including radiation production and CR acceleration, with the 
level of detail that is now achieved in 2D electron-positron shock 
simulations. 

The Weibel instability will also be studied in lab experiments at modern
Petawatt-scale laser facilities such as NIF, Omega EP, Hercules, PW, and 
Ghost. These lasers can produce relativistic electron beams with Lorentz 
factors of~$\sim 10^2$, similar to those in GRBs, and the Weibel 
turbulence has already been observed in some laser-plasma experiments
\cite{Tatarakis_etal-2003}. 
Therefore, the very possibility to experimentally explore the turbulence 
with good diagnostics is already at our disposal. Radiation mechanisms 
can be probed as well, since the predicted jitter radiation peaks at a 
few~keV and can be diagnosed with standard X-ray detectors. Amazingly, 
the laser-plasma parameters are very similar to those in GRB shocks, 
setting a characteristic length-scale (plasma skin depth) to be 
$\sim 10~\mu{\rm m}$ in laser plasma and a few mm in~GRBs. 
Apparently, not much scaling is needed between astrophysical 
phenomena and lab experiments. 
Producing a Weibel-mediated collisionless shock, which 
requires a few hundreds skin lengths to form, is clearly 
a feasible experiment with a typical cm-size target. 
Lighting up a GRB in a lab -- what can be more fascinating than this!





\begin{multicols}{2}

\renewcommand{\baselinestretch}{0.8} 

\small



\end{multicols}


\end{document}